\documentclass[conference, letterpaper]{IEEEtran}
\usepackage[backend=bibtex,style=numeric,sorting=none]{biblatex}

\usepackage[cmex10]{amsmath}
\usepackage{cleveref}
 
\usepackage{booktabs,threeparttable} 
\usepackage{subcaption}
\usepackage{tablefootnote}
\usepackage{todonotes}
\usepackage{amsfonts}
\footnoterule
\renewcommand{\footnoterule}{
  \kern -3pt                         
  \hrule width .5in
  \kern 2pt}  
\begin{document}
\title{Modeling and Contribution of Flexible Heating Systems for Transmission Grid Congestion Management}

\author{
\IEEEauthorblockN{David Kröger, Milijana Teodosic, Christian Rehtanz}
\IEEEauthorblockA{Institute of Energy Systems, Energy Efficiency and Energy Economics \\ 
TU Dortmund University, Germany\\
\{david.kroeger, milijana.teodosic, christian.rehtanz\}@tu-dortmund.de}
}

\maketitle

\thispagestyle{plain}
\pagestyle{plain}

\begin{abstract}
The large-scale integration of flexible heating systems in the European electricity market leads to a substantial increase of transportation requirements and consecutively grid congestions in the continental transmission grid. Novel model formulations for the grid-aware operation of both individual small-scale heat pumps and large-scale power-to-heat (PtH) units located in district heating networks are presented. The functionality of the models and the contribution of flexible heating systems for transmission grid congestion management is evaluated by running simulations for the target year 2035 for the German transmission grid. The findings show a decrease in annual conventional redispatch volumes and renewable energy sources (RES) curtailment resulting in cost savings of approximately 6\,\% through the integration of flexible heating systems in the grid congestion management scheme. The analysis suggests that especially large-scale PtH units in combination with thermal energy storages can contribute significantly to the alleviation of grid congestion and foster RES integration.
\end{abstract}

\vspace{0.618\baselineskip}


\section{Introduction}
The anticipated increase in cogeneration as well as the electrification of heat applications in Europe is a major building block in Europe's quest to decarbonize the energy sector \cite{EuropeanCommission.2016}. The accompanying stronger linkage between the thermal and the electricity sector results in various operational and planning challenges for the power system. For the coupled European electricity markets and continental transmission grid, one major challenge is the large-scale integration of power-to-heat (PtH) devices, namely heat pumps (HPs) and electric boilers (EBs), which represent an additional and partially flexible electrical load. On the supply side, the coupled generation from combined heat and power (CHP) plants mainly located in district heating networks (DHNs) - often in combination with large-scale thermal energy storages (TES) - is becoming more important as the phasing out of fossil fuels reduces the conventional generation capacity \cite{Lund.2015}. Currently, the European market design incentivizes the market-driven operation of PtH units which in turn leads to prospective high simultaneities and peak loads \cite{Kroger.2023}. Furthermore, due to the regional distribution of supply and demand, a substantial increase of transportation requirements and consecutively grid congestions can be expected from the increased sector coupling \cite{Xiong.2021}. While intra-zonal transmission grid congestion can basically occur in all bidding zones \cite{ENTSOE.2022}, the issues is expected to be predominant within the German bidding zone due to the growing regional discrepancy between load and supply which is fostered by the plans to strongly expand offshore wind capacities \cite{FederalMaritimeandHydrographicAgencyofGermany.2023}. Thus, effects on transmission grid congestion of alternative i.e. grid-aware operating schemes for sector-coupling technologies which take into account grid constraints are of high scientific interest. Although congestion is expected to be ubiquitous \cite{Metzger.2021}, there are yet few publications that model and quantify the impact of such grid-aware operating schemes in high detail within pan-European energy system analysis models. For the hydrogen sector, \cite{Xiong.2021} investigates the contribution of power-to-gas (PtG) units to alleviate transmission grid congestion. The study finds that by using flexibility from PtG in the electricity system, a reduction of renewable energy sources (RES) curtailment of 12\,\% can be achieved. \cite{Staudt.2018} assesses the ability of vehicle-to-grid as distributed storages to alleviate grid congestion and finds that electric vehicles (EVs) can contribute to transmission grid balancing. \cite{EmobRDMILES.2023} shows that the integration of flexible charging processes of battery EVs into congestion management can reduce market-driven simultaneities and the resulting grid congestions in the transmission grid. Focusing the heating sector, \cite{Schwaeppe.2022} applies a generation and transmission capacity expansion (GTCE) problem to investigate the flexibility provision from DHNs and decentral PtH units for transmission grid congestion. Due to the optimization design in which the capacities are endogenously determined, grid congestion is implicitly considered and prevented by design. The study finds that although DHNs offer strong cost reduction and flexibility potentials, the expansion of such networks does not significantly reduce the need for high-voltage grid expansion and should rather be understood as an element to provide short-term flexibility. A GTCE is also applied in \cite{Neumann.2023} to assess the benefits of a hydrogen network in Europe. Although playing a minor role, the study indicates that fuel cell CHP units can support transmission grid operation when grid expansion cannot be realized. 
A multi-stage sequential optimization that closer emulates the current market design is applied in \cite{Metzger.2021, Bertsch.2020}. \cite{Metzger.2021} highlights the expected ubiquity of grid congestion and stresses the contribution of flexibility-based congestion management from sector coupling technologies. Herein, the contribution of large-scale PtH devices for district heating and process heat is considered for congestion management while decentral PtH devices are considered inflexible. In \cite{Bertsch.2020} both decentral and large-scale PtH units are considered as flexibility options in the congestion management simulation and modeled as generic energy storages by introducing constraints that ensure balancing of thermal demand for each technology within a certain time period - every 3 hours for households and every 24 hours for district heating. The study indicates that the integration of flexibilites in the congestion management process has a positive, but comparably low influence on the overall required congestion management measures. 

Overall, the presented studies stress the growing issue of future congestion in the transmission grid and highlight the importance of a flexibility-based congestion management. However, in these studies the contribution of the heating sector for power grid congestion is either addressed implicitly as part of a GTCE problem formulation or in case of using a multi-stage sequential optimization only considered partially or with simplifications. This paper aims to contribute to the discussion by presenting novel model formulations for the grid-aware operation of both 1) individual small-scale building HPs and 2) large-scale PtH units located in DHNs. Subsequently, the contribution of those flexible heating systems for transmission grid congestion management is evaluated. The mathematical models are implemented in the pan-European energy system analysis model MILES \cite{MILES} and applied and validated to the future target year 2035 considering a high penetration of flexible PtH and CHP systems with focus on the German transmission grid.

\section{Methods}
\label{sec:methods}
The grid-aware integration of flexible heating systems is realized in the context of congestion management. The underlying redispatch problem is described in \ref{subsec:RD_objective}. The extension of the associated constraints is implemented for 1) electric-driven small-scale HPs as well as 2) large-scale PtH units in DHNs and is formulated in \ref{subsec:RD_extension}. The chapter ends with the definition of the system and grid-related contraints considering the (n-1)-criterion.

\subsection{Objective function}
\label{subsec:RD_objective}
The integration of flexible heating systems into the congestion management process requires a time-coupled approach to map the storage-like states. For this purpose, the objective function is formulated interval-based, allowing several time steps to be optimized simultaneously. An optimization interval describes the number of hours of a closed time period with $t^\text{start}$ being the first time step of each interval and $t^\text{inter}$ the associated interval duration. The objective is to minimize the costs of all congestion management measures of the entire interval \eqref{eq:RDZielfunktion}:

\begin{eqnarray} 
	\text{min}( \sum_{t=t^\text{start}}^{t^\text{start}+t^\text{inter}-1}  (\sum_{\theta \in \mathcal{M}_\theta} c_t^\theta + \sum_{\psi \in \mathcal{M}_\psi} c_t^\psi)), \quad \forall \: t_\text{start} \in \mathcal{M}_t^\text{start}
	\label{eq:RDZielfunktion}
\end{eqnarray}

The cost term is the sum of the underlying power adjustments of the traditional measures considered in congestion management \eqref{eq:RDtraditionelleKosten} as described in \cite{Spieker.62016} and the costs of additional flexibility potential resulting from the redispatch of flexible heating systems \eqref{eq:RDPtHKosten}.

\begin{eqnarray}
	\begin{split}
	c_t^\theta = c^+_{\theta,t}\cdot\Delta P^+_{\theta,t}+c^-_{\theta,t}\cdot\Delta P^-_{\theta,t} \quad \forall \: t \in \mathcal{M}_t
	\label{eq:RDtraditionelleKosten}
	\end{split}
\end{eqnarray}

\begin{eqnarray}
	\begin{split}
	c_t^\psi =(c^+_{\psi,t}\cdot\Delta P^+_{\psi,t}+c^-_{\psi,t}\cdot\Delta P^-_{\psi,t}) + \\
	 (c^{th, +}_{\psi,t}\cdot\Delta Q^{PE, +}_{\psi,t} + c^{th, -}_{\psi,t}\cdot\Delta Q^{PE, -}) \quad \forall \: t \in \mathcal{M}_t
	\label{eq:RDPtHKosten}
	\end{split}
\end{eqnarray}

The terms $c_t^\theta$ and $c_t^\psi$ describe the underlying costs for each control action in the respective time step. Regarding the case of flexible heating systems, the technology-specific costs $c_t^\psi$  result from the power adjustments in positive or negative direction $\Delta P^+_{\psi,t}$ or $\Delta P^-_{\psi,t}$ weighted with the corresponding costs for power increase $c^+_{\psi,t}$ or power decrease $c^-_{\psi,t}$. In case of CHP units, the adjustments of thermal power outputs $\Delta Q^{PE, +}_{\psi,t}$ and $\Delta Q^{PE, -}_{\psi,t}$ are considered taking into account the corresponding costs $c^{th, +}_{\psi,t}$ and $c^{th, -}_{\psi,t}$. The adjustments in the context of a power increase or reduction refer to the operating points that were determined in advance on the basis of a market simulation. The index $\psi$ describes an element of the set of all technology-specific units $\mathcal{M}_\psi$. $\mathcal{M}_\theta$ \eqref{eq:MengeTheta} and $\mathcal{M}_\psi$ \eqref{eq:MengePsi} represent the set of all considered technologies of the conventional congestion management and of flexible heating systems, respectively.  

\begin{eqnarray}
	\mathcal{M}_\Theta = \{ \text{HVDC, PST, PP} \not\ni \text{CHP, PS, RES, Dummy} \}
	\label{eq:MengeTheta}
\end{eqnarray}

\begin{eqnarray}
	\mathcal{M}_\Psi = \{ \text{HP, PtH, PP} \ni \text{CHP, TES, HOB} \}
	\label{eq:MengePsi}
\end{eqnarray}

 The conventional congestion management considers grid-related measures in the form of operating point adjustments of phase-shifting transformers (PST) and high-voltage direct current lines (HVDC), redispatch from conventional power plants (PP), pumped storages (PS) and curtailment of RES. To ensure the solvability of the optimization problem, fictive power reduction and power increase measures are integrated in the form of a slack variable. In the following, the consideration of each flexible heating system technology of the set $\mathcal{M}_\psi$ in congestion management and the underlying constraints are presented.

\subsection{Extension by flexible heating systems}
\label{subsec:RD_extension}

\subsubsection{Small-scale HPs}
\label{subsubsec:RD_i}
The flexibility of small-scale HPs is modeled by the thermal inertia of buildings. A 1RC1 building model is used to endogenously represent the heat demand of the units. The underlying building stock model is aggregated into archetypal building types $k_{BT}$ and regionalized into weather clusters $k_{WC}$. The clustering methods, as well as the determination of heat pump type dependent coefficient of power (COP) was presented in \cite{Kroger.2023}. In order for this approach to be transferred to congestion management and for the small-scale HPs to be used in a grid-aware way, the spatial resolution in terms of weather clusters  is disaggregated and mapped equivalently to grid nodes $n$. For each weather cluster $WC \in \mathcal{M}_{WC}$ the grid node $n_{WC} \in \mathcal{M}_n$ is identified which has the smallest distance $\delta^{r,n}$ to the region $r_{WC} \in \mathcal{M}_r$ with $\lambda$ and $\mu$ representing the longitudes and latitudes of regions and nodes \eqref{eq:DistanzKnoten}. The underlying coefficient of power (COP) for the considered HP technologies as well as the occurrences of the building types per weather cluster are transferred analogously to node level.

\begin{eqnarray} 
	\begin{split}
	\delta^{r,n}=\sqrt{(\lambda_r-\lambda_n)^2+(\mu_r-\mu_n)^2} \\\ \quad \forall \: r \in \mathcal{M}_r, \: \forall \: n \in \mathcal{M}_n
	\label{eq:DistanzKnoten}
	\end{split}
\end{eqnarray}

In order to reduce the complexity of the optimization problem which is necessary to enable solving in an acceptable time, an aggregation of building types on nodal level is conducted. Therefore, one virtual building per node that represents the building types and their occurrence is established by interconnecting the individual R and C values of all buildings considering their occurrence \eqref{eq:Aggregation_R} - \eqref{eq:Aggregation_C}. Here, $nb_{BT,n}$ represents the occurrence of each building type per node.

\begin{eqnarray} 
		\frac{1}{R_n} = \sum_{BT \in \mathcal{M}_{BT}} \frac{nb_{BT,n}}{R_{BT}} \quad \forall \: n \in \mathcal{M}_{n}
		\label{eq:Aggregation_R}
\end{eqnarray}
\begin{eqnarray} 
	C_n = \sum_{BT \in \mathcal{M}_{BT}} nb_{BT,n} \cdot {C_{BT}} \quad \forall \: n \in \mathcal{M}_{n}
	\label{eq:Aggregation_C}
\end{eqnarray}

The impact of the thermal inertia is represented by permitting the indoor temperature to fluctuate within an upper and lower bound. For this purpose, a temperature range is introduced for the heating system. Due to the optimization formulation as a delta consideration, constant heat sources such as solar radiation or heat flux from persons does not have to be explicitly modeled in the congestion management, because their influence is implicitly considered within the permissible range of temperature and delta heating power resulting from the market optimization. The change in indoor temperature at each node in the corresponding time step is described by equation \eqref{eg:DeltaTIndoor}. $\Delta \Phi^{\mathrm{Heating}}_{t,n}$ describes the change in heating power compared to the base heating power from the market dispatch. The maximum heating power is restricted by \eqref{eq:DeltaPhiLimits}. The change in indoor temperature plus the prior determined indoor temperature $T^{\mathrm{In,market}}_{t,n}$ has to be within the permissible upper and lower bounds \eqref{eq:DeltaTLimits}. 

\begin{eqnarray}
	\begin{split}
		\Delta T^{\mathrm{In}}_{t,n} = \Delta T^{\mathrm{In}}_{t-1,n} - \frac{1}{R_{n}C_{n}} (\Delta T^{\mathrm{In}}_{t-1,n}) + \frac{1}{C_{n}} (\Delta \Phi^{\mathrm{Heating}}_{t,n})
		\\
		 \quad \forall \: t \in \mathcal{M}_{t} \setminus \mathcal{M}_{t_{\mathrm{start}}}, \: n \in \mathcal{M}_n		 
	\label{eg:DeltaTIndoor}
	\end{split}
\end{eqnarray}

	\begin{eqnarray}
		\begin{split} 
 	 T^{\mathrm{In,min}}_{t,n} \leq T^{\mathrm{In,market}}_{t,n} + \Delta T^{\mathrm{In}}_{t,n} \leq T^{\mathrm{In,max}}_{t,n}
	\\
		\quad \forall \: t \in \mathcal{M}_{t} \setminus \mathcal{M}_{t_{\mathrm{start}}}, \: n \in \mathcal{M}_n
		\label{eq:DeltaTLimits}
		\end{split}	
	\end{eqnarray}

	\begin{eqnarray}
		\begin{split} 
 	0 \leq \Phi^{\mathrm{Heating,market}}_{t,n} + \Delta \Phi^{\mathrm{Heating}}_{t,n} \leq \Phi^{\mathrm{Heating,max}}_{t,n}
	\\
		\quad \forall \: t \in \mathcal{M}_{t} \setminus \mathcal{M}_{t_{\mathrm{start}}}, \: n \in \mathcal{M}_n
		\label{eq:DeltaPhiLimits}
		\end{split}	
	\end{eqnarray}

The transformation to a change of the electrical load $\Delta P^{\mathrm{Heating}}_{t,n}$ per node is performed by means of the COP \eqref{eq:DeltaPHeating}. The decision variables included in the objective function are formulated in equation \eqref{eq:DeltaPHP+} and \eqref{eq:DeltaPHP-}, and each describes the aggregated adjusted electrical load for all nodes. $\Delta P^{\mathrm{HP,+}}_{\psi,t}$ results from the increase of the heating power and $\Delta P^{\mathrm{HP,-}}_{\psi,t}$ from a decrease. When considering a cooling system, the decision variables and constraints can be formulated vice versa to the presented approach.

\begin{eqnarray}
	\begin{split}
	\Delta P^{\mathrm{Heating}}_{t,n} = \Delta \Phi^{\mathrm{Heating}}_{t,n} \cdot \frac{1}{COP_{t,n}} \\
		\quad \forall \: t \in \mathcal{M}_{t}, \: n \in \mathcal{M}_n
	\label{eq:DeltaPHeating}
	\end{split}
\end{eqnarray}

\begin{subequations}
\begin{eqnarray}
	\begin{split}
	\Delta P^{\mathrm{HP,+}}_{\psi,t} = \sum_{n \in \mathcal{M}_n} \sum_{BT \in \mathcal{M}_{BT}} \Delta P^{\mathrm{Heating}}_{t,BT,n}  \\\
	\quad \forall \:  \Delta P^{\mathrm{Heating}}_{t,BT,n} \in \mathbb{R}^{\geq 0},  \: t \in \mathcal{M}_{t}, \: BT \in \mathcal{M}_{BT}, \: n \in \mathcal{M}_n
	\label{eq:DeltaPHP+} 	
	\end{split} 
	\\
	\begin{split}
		\Delta P^{\mathrm{HP,-}}_{\psi,t} = \sum_{n \in \mathcal{M}_n} \sum_{BT \in \mathcal{M}_{BT}} 
	\Delta P^{\mathrm{Heating}}_{t,BT,n} \\\
	\quad \forall \:  \Delta P^{\mathrm{Heating}}_{t,BT,n} \in \mathbb{R}^{\leq 0}, \: t \in \mathcal{M}_{t}, \: BT \in \mathcal{M}_{BT}, \: n \in \mathcal{M}_n
	\label{eq:DeltaPHP-}
	\end{split}
\end{eqnarray}
\end{subequations} \\
	 
\subsubsection{Large-scale PtH units}
\label{subsubsec:RD_ii}
Both (ii) large-scale PtH and CHP units in DHNs are also mapped to grid nodes within the congestion management optimization to create a link between their operation and the resulting grid load. Through this modeling approach, regionally different dispatch results of PtH and CHP units in combination with large-scale TES and gas-fired peak-load boilers can be determined for each DHN considered. 

Equation \eqref{eq:Delta_DHN} ensures the heat balance for each considered DHN after grid-related adjustments/changes in the heat output of thermal conversion units, namely thermal power plants $\dot{Q}^{\mathrm{CHP}}_{\psi, t}$, PtH units $\dot{Q}^{\mathrm{PtH}}_{\psi, t}$, heat-only boilers (HOB) $\dot{Q}^{\mathrm{HOB}}_{\psi, t}$ and TES $\dot{Q}^{\mathrm{TES}}_{\psi, t}$. Please note that by using this formulation the individual terms can take both positive and negative values. In case of TES, a positive sign means charging and a negative sign discharging.

\begin{eqnarray}
	\begin{split}
		0 = \sum_{\psi \in \mathcal{M}_{CHP,i}} \Delta \dot{Q}^{\mathrm{CHP}}_{\psi, t} + \sum_{\psi \in \mathcal{M}_{PtH,i}} \Delta \dot{Q}^{\mathrm{PtH}}_{\psi, t} \\
		+ \sum_{\psi \in \mathcal{M}_{HOB,i}} \Delta \dot{Q}^{\mathrm{HOB}}_{\psi, t}
		- \sum_{\psi \in \mathcal{M}_{TES,i}} \Delta \dot{Q}^{\mathrm{TES}}_{\psi, t} \\
		\quad \forall \: i \in \mathcal{M}_{DHN}, \: t \in \mathcal{M}_t
		\label{eq:Delta_DHN}
	\end{split}
\end{eqnarray}

The operating point of a CHP plant can be described by means of its power and thermal output and the corresponding primary fuel consumption. Thus, \eqref{eq:Q_PE_BKP} and \eqref{eq:Q_PE_EXT} establish a relationship between the grid-related adjustments of power and thermal output and the resulting change in fuel consumption for back-pressure turbines (BKP) and extraction-condensing turbines (EXT), respectively. 

\begin{eqnarray}
	\begin{split}
		\Delta Q^{\mathrm{PE}}_{\psi,t} = \frac{1}{\eta_\psi^{\mathrm{cond}}} \cdot \Delta P^{\mathrm{CHP}}_{\psi, t} \\
		\forall \: \psi \in \mathcal{M}_{BKP}, \: t \in \mathcal{M}_t
		\label{eq:Q_PE_BKP}
	\end{split}
\end{eqnarray}

\begin{eqnarray}
	\begin{split}
		\Delta Q^{\mathrm{PE}}_{\psi,t} = \frac{1}{\eta_\psi^{\mathrm{cond}}} \cdot \Delta P^{\mathrm{cond}}_{\psi, t} 
		+ \frac{1 + \sigma_\psi}{\eta_\psi^{\mathrm{total,max}}} \cdot \Delta \dot{Q}^{\mathrm{CHP}}_{\psi, t} \\
		\forall \: \psi \in \mathcal{M}_{EXT}, \: t \in \mathcal{M}_t
		\label{eq:Q_PE_EXT}
	\end{split}
\end{eqnarray}

In case of PtH units, the change in power consumption is linked to the change of the heat output via the (present) COP \eqref{eq:Q_Delta_PtH}. Furthermore, the change of power consumption added to the predetermined market-based operating point has to be within the feasible technical limits \eqref{eq:P_Delta_PtH}. 
The same applies analogous for the change of thermal output and the predetermined market-based thermal output of TES \eqref{eq:Q_Delta_TES} and HOB \eqref{eq:Q_Delta_HOB}.
TES are described by means of a linear system equation that incorporates level-depended losses \eqref{eq:TES_Zustand}. Due to the implementation of the congestion management simulation with no coupling between the single optimization intervals in order to enable parallel solving, the initial and end states of the TES are set to the (predetermined) market-based state of charges \eqref{eq:TES_Anfang} - \eqref{eq:TES_End}.

\begin{eqnarray}
		\Delta \dot{Q}^{\mathrm{PtH}}_{\psi, t}  = \Delta P^{\mathrm{PtH}}_{\psi, t} \cdot COP_{\psi, t}^{\mathrm{PtH}}
		\quad \forall \: \psi \in \mathcal{M}_{PtH}, \: t \in \mathcal{M}_t
		\label{eq:Q_Delta_PtH}
\end{eqnarray}

\begin{eqnarray}
		\begin{split}
		0 \leq P^{\mathrm{PtH,market}}_{\psi, t} + \Delta P^{\mathrm{PtH}}_{\psi, t} \leq P^{\mathrm{PtH,max}}_{\psi} \\
		\forall \: \psi \in \mathcal{M}_{PtH}, \: t \in \mathcal{M}_t
		\label{eq:P_Delta_PtH}
		\end{split}
\end{eqnarray}

\begin{eqnarray}
	\begin{split}
		E^{\mathrm{TES}}_{\psi, t} = E^{\mathrm{TES}}_{\psi, t-1} \cdot \eta^{\mathrm{TES}}_{\psi} + (\dot{Q}^{\mathrm{TES, market}}_{\psi, t} + \Delta \dot{Q}^{\mathrm{TES}}_{\psi, t}) \\
		\forall \: \psi \in \mathcal{M}_{TES}, \: t \in \mathcal{M}_t \setminus \mathcal{M}_{t_{\mathrm{start}}}
		\label{eq:TES_Zustand}
	\end{split}
\end{eqnarray}

\begin{eqnarray}
		\begin{split}
		0 \leq \lvert \dot{Q}^{\mathrm{TES,market}}_{\psi, t} + \Delta \dot{Q}^{\mathrm{TES}}_{\psi, t} \rvert \leq \dot{Q}^{\mathrm{TES, max}}_{\psi} \\
		\forall \: \psi \in \mathcal{M}_{TES}, \: t \in \mathcal{M}_t
		\label{eq:Q_Delta_TES}
		\end{split}
\end{eqnarray}

\begin{eqnarray}
		\begin{split}
		0 \leq E^{\mathrm{TES}}_{\psi, t} \leq E^{\mathrm{TES,max}}_{\psi} \\
		\forall \: \psi \in \mathcal{M}_{TES}, \: t \in \mathcal{M}_t
		\label{eq:TES_minMax}
		\end{split}
\end{eqnarray}

\begin{eqnarray}
	\begin{split}
		E^{\mathrm{TES}}_{\psi, t} &= E^{\mathrm{TES,init,market}}_{\psi} \\
		&\forall \: \psi \in \mathcal{M}_{TES}, \: t = t^{\mathrm{start}}
		\label{eq:TES_Anfang}
	\end{split}
\end{eqnarray}

\begin{eqnarray}
	\begin{split}
		E^{\mathrm{TES}}_{\psi, t} &= E^{\mathrm{TES,end,market}}_{\psi} \\
	&\forall \: \psi \in \mathcal{M}_{TES}, \: t = t^{\mathrm{start}} + t^{\mathrm{inter}} - 1
\label{eq:TES_End}
	\end{split}
\end{eqnarray}

\begin{eqnarray}
		\begin{split}
		0 \leq \dot{Q}^{\mathrm{HOB,market}}_{\psi, t} + \Delta \dot{Q}^{\mathrm{HOB}}_{\psi, t} \leq \dot{Q}^{\mathrm{HOB, max}}_{\psi} \\
		\forall \: \psi \in \mathcal{M}_{HOB}, \: t \in \mathcal{M}_t
		\label{eq:Q_Delta_HOB}
		\end{split}
\end{eqnarray}

\subsection{System and grid-related constraints}
In addition to the technology-specific constraints for the flexible heating systems, technology-wide constraints responsible for ensuring power balances and grid constraints are introduced. The total power adjustments of all technologies must be balanced at each time step \eqref{eq:PowerBalance}. The adjustment per unit results from the superposition of positive and negative actions of the corresponding unit \eqref{eq:DeltaPTheta} - \eqref{eq:DeltaPPsi}.  

\begin{eqnarray} 
	\sum_{\theta \in \mathcal{M}_\theta} \Delta P_{\theta,t} + \sum_{\psi \in \mathcal{M}_\psi} \Delta P_{\psi,t} = 0 \quad \forall \: t \in \mathcal{M}_t
	\label{eq:PowerBalance}
\end{eqnarray}  

\begin{eqnarray} 
	\begin{split}
		\Delta P_{\theta,t} = \Delta P^+_{\theta,t} - \Delta P^-_{\theta,t} \\\
		\quad \forall \: \theta \in \mathcal{M}_\theta, \forall \: t \in \mathcal{M}_t
		\label{eq:DeltaPTheta}
	\end{split}
\end{eqnarray} 

\begin{eqnarray} 
	\begin{split}
		\Delta P_{\psi,t} = \Delta P^+_{\psi,t} - \Delta P^-_{\psi,t} \\\
		\quad \forall \: \psi \in \mathcal{M}_\psi, \forall \: t \in \mathcal{M}_t
		\label{eq:DeltaPPsi}
	\end{split}
\end{eqnarray} 

Equation \eqref{eq:PLimitN0} ensures that congestions are resolved in the undisturbed case for all grid elements from the set $\mathcal{M}_l$ intendend to be protected by congestion management.

\begin{eqnarray} 
	\begin{split}
		-P^\text{max}_{l,t} \leq P_{l,t}^\text{(n-0)} \leq P^\text{max}_{l,t} \\\
		\quad \forall \: t \in \mathcal{M}_t, \forall \: l \in \mathcal{M}_l
		\label{eq:PLimitN0}
	\end{split}
\end{eqnarray} 

The power flow in the undisturbed case $P_{l,t}^\text{(n-0)}$ is determined by the initial flow $P_{l,t}^\text{init}$ prior congestion management modified by the changes resulting from the congestion management measures (\ref{eq:PFlowN0}). 

\begin{eqnarray}
	\begin{split}
		P_{l,t}^\text{(n-0)} = 	P_{l,t}^\text{init} + \sum_{n \in \mathcal{M}_n} PTDF_{n \rightarrow l} \cdot \Delta P^\text{RD}_{n,t} + ...\\\
		... \sum_{n \in \mathcal{M}_n} PSDF_{n \rightarrow l} \cdot (\Delta \alpha^+_{n,t} - \Delta \alpha^-_{n,t}) \\\
		\quad \forall \: t \in \mathcal{M}_t, \forall \: l \in \mathcal{M}_l
		\label{eq:PFlowN0}
\end{split}
\end{eqnarray}

The aggregated node power $\Delta P^\text{RD}_{n,t}$ after congestion management is the sum of the operating point adjustments of all considered units $\mathcal{M}_\theta(n)$ or $\mathcal{M}_\psi(n)$ at node $n$ \eqref{eq:DeltaNode}.

\begin{eqnarray} 
	\begin{split}
		\Delta P^\text{RD}_{n,t} = \sum_{\theta \in \mathcal{M}_\theta(n)} \Delta P_{\theta,t} + \sum_{\psi \in \mathcal{M}_\psi(n)} \Delta P_{\psi,t} \\\
		\quad \forall \: t \in \mathcal{M}_t, \forall \: n \in \mathcal{M}_n 
		\label{eq:DeltaNode}
	\end{split}
\end{eqnarray} 

To transfer the influence of the power changes at the nodes to a line $l$, these are multiplied by the underlying linear sensitivities \textit{Power Transfer Distribution Factors} $PTDF_{n \rightarrow l}$ \eqref{eq:PTDF}. Equivalently, the \textit{Phase Shift Distribution Factors} $PSDF_{n \rightarrow l}$ map the influence of a phase change $\Delta \alpha^+_{n,t}$ or $\Delta \alpha^-_{n,t}$ of a PST at node $n$ on the resulting power flow (\ref{eq:PSDF}).

\begin{eqnarray} 
\begin{split}
	PTDF_{n \rightarrow l} = \frac{\Delta P_l}{\Delta P_n} \\\
	 \quad \forall \: l \in \mathcal{M}_l, \forall \: n \in \mathcal{M}_n 
	\label{eq:PTDF}
\end{split}
\end{eqnarray} 

\begin{eqnarray} 
	\begin{split}
		PSDF_{n \rightarrow l} = \frac{\Delta P_l}{(\Delta \alpha^+_{n,t} - \Delta \alpha^-_{n,t})} \\\
		\quad \forall \: l \in \mathcal{M}_l, \forall \: n \in \mathcal{M}_n
		\label{eq:PSDF}
	\end{split}
\end{eqnarray} 

In addition to the undisturbed case, the grid elements are also protected in the case of a failure. The (n-1)-criterion is considered for a possible contingency (C) of a line and represented by linear \textit{Line Outage Distribution Factors} $LODF_{C \rightarrow l}$ \eqref{eq:LODF}. The influence of the contingency of  line C is transferred to the power flows of the considered grid element $l$ \eqref{eq:PLimitN1}.

\begin{eqnarray}
\begin{split}
-P^\text{max}_{l,t} \leq P_{C,l,t}^\text{(n-1)} \leq P^\text{max}_{l,t} \\\
	\quad \forall \: t \in \mathcal{M}_t, \forall \: l \in \mathcal{M}_l, \forall \: C \in \mathcal{M}_{C}
\label{eq:PLimitN1}
\end{split}
\end{eqnarray} 

\begin{eqnarray}
\begin{split}
P_{C,l,t}^\text{(n-1)} &= P_{l,t}^\text{(n-0)} + LODF_{C \rightarrow l} \cdot P_{C,t}^\text{(n-0)} \\\ 
&\forall \: t \in \mathcal{M}_t, \forall \: l \in \mathcal{M}_l, \forall \: C \in \mathcal{M}_{C}
\label{eq:LODF} 
\end{split}
\end{eqnarray}

\section{Scenario and Results}
To test the functionality of the models and assess the contribution of flexible heating systems for transmission grid congestion management, simulations for the target year 2035 are carried out and evaluated. The scenario is based on the federal German network development plan (2021, \cite{NEP.2021}) and the \textit{Distributed Energy} scenario from the ten year network development plan (TYNDP) (2020, \cite{TYNDP.2020}). Key figures of generation and flexible heating capacities for the German system are listed in Table \ref{tab:ScenarioDataGermany}. The underlying grid model refers to the interconnected system of continental Europe. The grid expansion status for Germany is based on \cite{NEP.2021},  which is characterized by a considerable expansion of HVDCs along the corridor from north to south. The pan-European transmssion grid consists of approximately 4,000 nodes and 6,000 branches. The extended congestion management simulation is focused on the German transmission grid, which is represented by approximately 680 nodes and 800 branches.   

\begin{table}\centering 
	\caption{Scenario data Germany} \label{tab:ScenarioDataGermany}
	\begin{tabular}{@{}llc@{}}
		\toprule
		Parameter & Unit & 2035 \\
		\midrule
		Generation capacities && \\
		\quad Natural gas & GW & 48.2  \\ 
		\quad Onshore wind & GW & 86.8 \\
		\quad Offshore wind & GW & 30 \\
		\quad Photovoltaics & GW & 117.8 \\
		Flexible heating systems && \\
		\quad S-s\tablefootnote{Small-scale}  HPs & GW & 39  \\ 
		\quad L-s\tablefootnote{Large-scale}  HPs & GW & 4  \\ 
		\quad L-s EBs & GW & 2  \\	
		\quad TES in DHN & GWh & 80  \\	
		\bottomrule
	\end{tabular}
\end{table}

The experimental setup consists of two simulation runs that differ in terms of flexibility provision from heating systems. In both cases, all electricity-based heating systems are considered flexible within their respective bounds and have the technical requirements necessary to react to price or grid signals. In the base case (base), flexibility from heating systems is only provided to the electricity market and the congestion management scheme is carried out without adjusting the operating points of heating systems. In the comparative case (flex), flexibility is first provided to the market, but now the operating points of flexible heating systems can be adjusted within the congestion management resulting in additional degrees of freedom for the optimization. The models and simulations have been implemented and carried out using MATLAB ver. R2022a in combination with the toolbox for optimization modeling YALMIP and Gurobi Optimizer 10.0.3 \cite{TheMathWorksInc..2022, Lofberg2004, gurobi}. Generally, the simulations are carried out in hourly resolution with an interval length of 168 hours. Although, due to the complexity of the time-coupled simulation and to achieve reasonable computation times, it was necessary to decompose the optimization in the flex case to 24 hours intervals to determine the operating points of small-scale HPs. These operating points are then used as inputs for an optimization with an interval length of 168 hours. The simulations have been carried out on a workstation using 16 cores on AMD EPYC 7502P with 256 GB RAM. The original formulation considering small-scale HPs and an interval length of 168 hours did not find a feasible solution within 72 hours. The decomposed optimization with small-scale HPs provided results in approximately 6 hours (24 hours interval) and 4 hours (168 hours interval).    

\begin{table}\centering 
	\caption{Electric and thermal redispatch volumes} \label{tab:Redispatchmengen}
	\begin{tabular}{@{}lccclccc@{}}
		\toprule
		$\mathrm{TWh_{el}}$ && base & flex & $\mathrm{TWh_{th}}$ && base & flex \\
		\midrule
		RES & (-) & 24.24 & 23.53 & RES & (-) & 0 & 2.16  \\ 
		PP and dummy & (-) & 15.16 & 14.65 & CHP & (-) & 8.31 & 9.56  \\
		S-s HPs & (+) & 0 & 2.31 & L-s PtH & (-) & 0 & 9.97 \\
		L-s PtH & (+) & 0 & 3.22 & HOB & (-) & 3.59 & 2.58 \\ [5pt]
		PP and dummy & (+) & 39.4 & 37.11 & RES & (+) &  2.16 & 2.77 \\
		S-s HPs & (-) & 0 & 2.31 &  CHP & (+) & 7.56 & 10.72 \\
		L-s PtH & (-) & 0 & 4.29 & L-s PtH & (+) & 0 & 6.73  \\
		&&&& HOB & (+) & 2.18 & 4.05  \\	
		\bottomrule
	\end{tabular}
\end{table}

\begin{figure*}
	\centering
	\includegraphics[width=\textwidth]{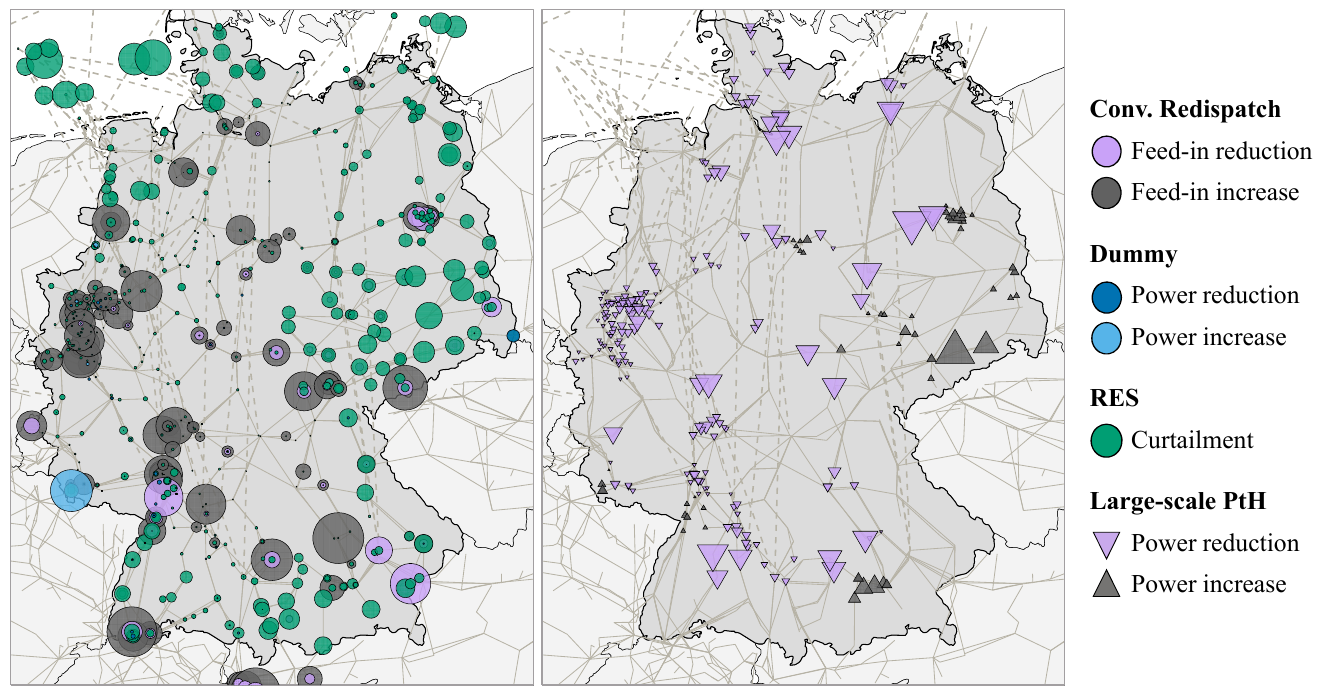}
	\caption{Left: Annual redispatch volumes in the elctricity domain for the flex case. Right: Annual change of PtH consumption for the flex case.} 
	\label{fig:ErgNetzDE} 
\end{figure*}

Table \ref{tab:Redispatchmengen} shows the electrical and thermal redispatch volumes for the base and flex case. In both the electric and thermal domain, considerable amounts of positive and negative power adjustments have been carried out in the congestion management simulation. The integration of flexible heating systems leads to a minor, but significant decrease in conventional redispatch and RES curtailment volumes in the elctricity domain. The amount of RES curtailment is reduced by 0.71\,TWh indicating the positive effect of flexible heating systems for the integration of RES. Furthermore, negative and positive conventional redispatch volume is decreased by 0.51\,TWh and 2.29\,TWh, respectively. The reduction of the conventional redispatch and RES curtailment volumes lead to a cost saving regarding the total variable system cost of approximately 6\,\% between the flex and base case. Interestingly, the amount of positive and negative large-scale PtH volumes do not balance each other out meaning that in the flex case an overall curtailment of PtH units has been carried out. This is enabled by the potential alternative heat provision in DHNs via thermal RES, CHP plants and HOB. Thus, the PtH curtailment propagates to the thermal domain resulting in thermal redispatch volumes. The curtailment of 2.83\,TWh thermal feed-in from PtH units is primarily compensated by HOBs ($+ 1.47 \: \mathrm{TWh}$) and CHP units ($+ 1.16 \: \mathrm{TWh}$). Considering the spatial distribution of thermal redispatch volumes (Figure \ref{fig:ErgNetzDE}), two effects can be observed. (i) A decrease in large-scale PtH consumption frequently occurs in conjunction with an increase of thermal power output from CHP units resulting from the conventional electricity redispatch. (ii) An increase in PtH consumption frequently occurs in regions with RES surplus as can be seen for example in the eastern and south-eastern parts of Germany.  

\begin{figure}
	\centering
	\includegraphics[width=0.5\textwidth]{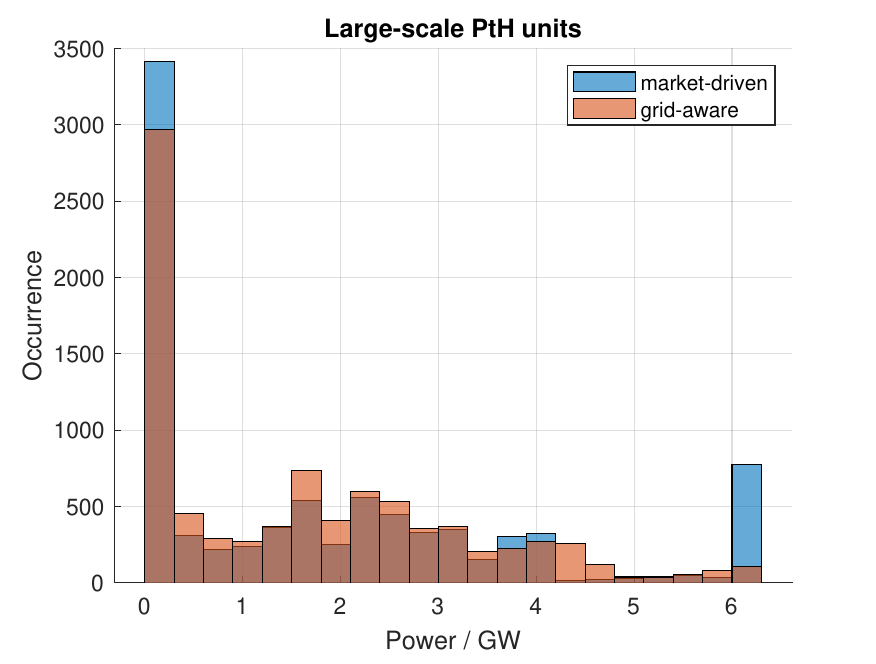}
	\caption{Distribution of operating points of large-scale PtH units in DHN.} 
	\label{fig:histLargeScalePtH} 
\end{figure}

In addition to the change of spatial power consumption of PtH units, a change in the temporal distribution can also be observed. Figure \ref{fig:histLargeScalePtH} depicts the frequency of operating points of large-scale PtH units for the market-driven and grid-aware operation scheme. In the grid-aware scheme, a strong decrease of full-load (6\,GW) operation hours can be observed indicating the grid-related curtailment of maximum power consumption. Furthermore, the total decrease of 1.07\,$\mathrm{TWh_{el}}$ electricity consumption from large-scale PtH (see Table \ref{tab:Redispatchmengen}) is noticeable.  

\begin{figure}
	\centering
	\includegraphics[width=0.5\textwidth]{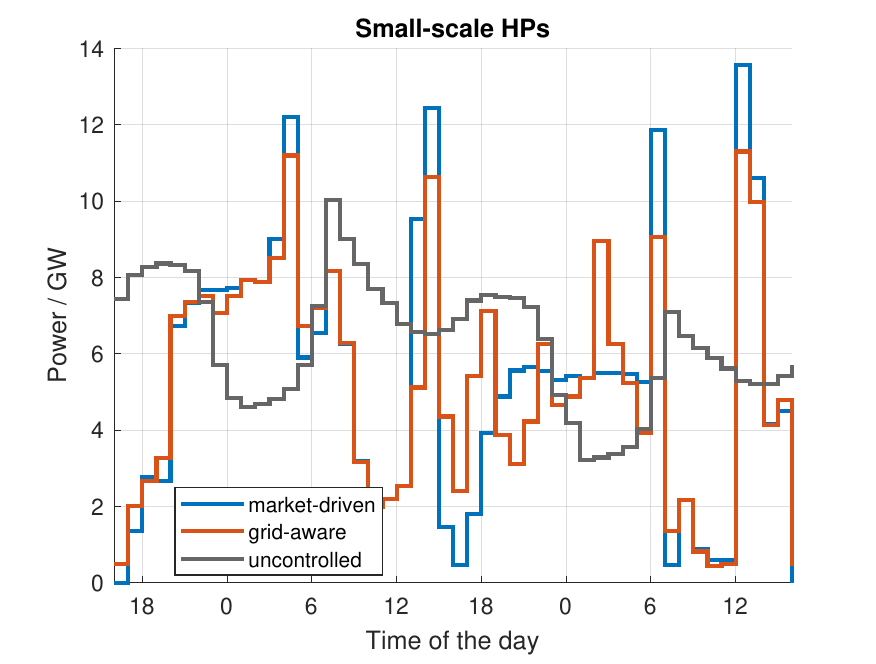}
	\caption{Snapshot of uncontrolled, market-driven and grid-aware operating scheme of small-scale HPs for two days.} 
	\label{fig:Operation_smallScaleHP} 
\end{figure}

In case of small-scale HPs, no alternative heating source at the house level is available. Thus, the grid-aware operating scheme yields no change in the regional distribution of annual consumption figures. Nevertheless, small-scale HPs can offer short-term flexibility which is illustrated by a change in their temporal consumption pattern. Figure \ref{fig:Operation_smallScaleHP} illustrates three different operating strategies for small-scale HPs, namely uncontrolled, market-driven and grid-aware. The uncontrolled operation has been determined based on a standard load profile approach and exhibits a comparatively uniform characteristic. In case of market-driven operation, a high simultaneity can be observed during times with high RES supply or low electrical demand that occurs typically during midday or in the early morning. The grid-aware operation shows a similar pattern, but with reduced peaks in order to reduce grid congestion.  

\begin{figure}
	\centering
	\includegraphics[width=0.5\textwidth]{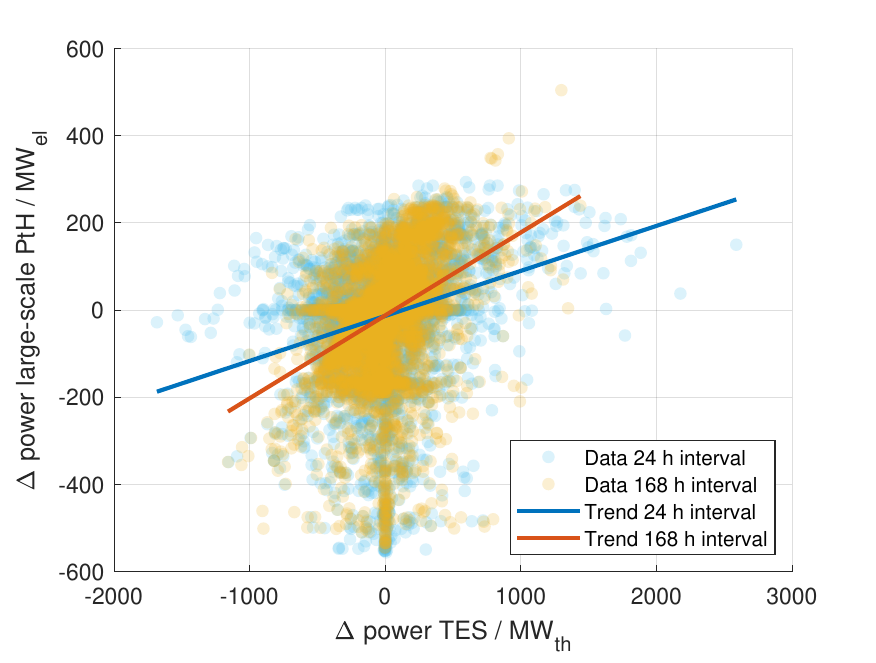}
	\caption{Correlation of change of operating points for large-scale PtH and TES.} 
	\label{fig:PtHvsTES} 
\end{figure}

Flexibility in DHNs is enabled to a large extent by the integration of large-scale TES. Figure \ref{fig:PtHvsTES} illustrates the positive correlation between an increase in PtH consumption and an increase in TES charging and vice versa for the 24 and 168 hours optimization interval. This indicates that TES in combination with PtH are a suitable option to virtually store electricity in the heating domain. Furthermore, the stronger correlation in the 168 hours optimization interval shows that the interval length has a strong influence on the operation of PtH und TES units in DHNs and suggests that TES are a suitable long-term storage option for RES. 

\section{Discussion and Conclusion}
In what follows, a few caveats of this study will be discussed. First, it has been presumed that all small-scale HPs have the technical equipment and the willingness of the users to be operated in a flexible manner. Considering the additional costs for the equipment necessary and potential comfort losses resulting from the flexibility provision, the overall practicability has to be scrutinized. Furthermore, no restrictions from the distribution system have been considered in this study. Taking into account the high simultaneities of small-scale HPs, a prior distribution-system-related curtailment of power is conceivable. These limitations apply mainly to small-scale HPs and to a lesser extent to industrial-size units which can be connected to higher voltage levels and which are typically operated by corporations stronger driven by monetary incentives than individual households. Nevertheless, the presented models are formulated in a general way and can also be applied to distribution system congestion management. The spatial scope of the congestion management case study has been the German transmission grid which in future works should be extended to also include other European countries with their particular generation and load composition to evaluate the transferability of the findings. The evaluations have been carried out using the multi-stage sequential optimization model MILES that emulates the European market design. MILES is a fundamental simulation framework that does not consider investment costs for flexibility provision. The results of the optimization show a strong sensitivity regarding the interval length of the optimization. In our case, the length has been set to 168 hours to guarantee solving within an acceptable time period. Future work should include the development of adequate procedures to reduce the complexity of the optimization in order to better assess the long-term storage capabilities of large-scale TES and other storage types.

Overall, the results confirm the functionality of the developed models and show the importance of an adequate representation of the heating sector for transmission grid congestion management. The distinction between individual small-scale building HPs and large-scale PtH units located in DHNs enables a profound analysis and evaluation of the contribution of flexible heating systems for transmission grid congestion management. In case of small-scale HPs, a grid-aware operation can reduce the resulting peak loads and contribute to grid alleviation, but due to high simultaneities during cold periods, relative high losses and small storage volumes, their flexibility provision with regards to annual redispatch volumes can be considered small. In contrast, while representing only a fraction of the total PtH capacity - 6\,GW out of a total of 45\,GW - large-scale PtH units have a stronger impact on the transmission grid congestion management. This can mainly be attributed to three characteristics: PtH units in DHNs can be coupled with large-scale TES (i) offering significant storage capacities while (ii) maintaining comparatively low losses. Furthermore, they can generally be operated with high flexibility due to (iii) opportunities resulting from diverse asset portfolio in DHNs including CHP plants which pose an important complementary part. The interaction between the electricity and the heating sector culminates in corresponding redispatch volumes in the heating domain. 

These findings are in line with the current state of affairs and illustrate the potential of DHNs for the integration of RES and transmission grid congestion management. Thus, the presented models and findings can be a starting point for further research considering different target years and scenarios. Two intriguing questions would be the evaluation of grid expansion measures considering the flexibility of heating systems and the determination of an optimal capacity allocation of units in DHNs to optimize the overall energy system.
 
\section*{Acknowledgments} 
This research was partly funded by the Ministry of Culture and Science of the State of North Rhine-Westphalia within the Graduate School for \textit{Sustainable Energy Systems in the Neighbourhoods} and the BMWK within the project GreenVEgaS (funding code: 03EI1009A). 


\printbibliography

\end{document}